\documentclass{article}
\usepackage{ismir,amsmath,amssymb,cite}
\usepackage{graphicx}
\usepackage{subcaption}
\usepackage{color}
\usepackage{url}

\sixauthors
{Rainer Kelz}
{Matthias Dorfer}
{Filip Korzeniowski}
{Sebastian B\"{o}ck}
{Andreas Arzt}
{Gerhard Widmer}
{Department of Computational Perception, Johannes Kepler University Linz, Austria}
{\texttt{rainer.kelz@jku.at}}

\title{On the Potential of Simple Framewise Approaches to Piano Transcription}

\newcommand{\x}{$\times$}

\begin{document}

\maketitle

\begin{abstract}
In an attempt at exploring the limitations of simple approaches to the task of piano transcription (as usually defined in MIR), we conduct an in-depth analysis of neural network-based framewise transcription. We systematically compare different popular input representations for transcription systems to determine the ones most suitable for use with neural networks. Exploiting recent advances in training techniques and new regularizers, and taking into account hyper-parameter tuning, we show that it is possible, by simple bottom-up frame-wise processing, to obtain a piano transcriber that outperforms the current published state of the art on the  publicly available MAPS dataset -- without any complex post-processing steps. Thus, we propose this simple approach as a new baseline for this dataset, for future transcription research to build on and improve.
\end{abstract}

\section{Introduction}
\label{sec:introduction}
Since their tremendous success in computer vision in recent years, neural networks have been used for a large variety of tasks in the audio, speech and music domain. They often achieve higher performance than hand-crafted feature extraction and classification pipelines \cite{lecun_deeplearning}. Unfortunately, using this model class brings along considerable computational baggage in the form of hyper-parameter tuning. These hyper-parameters include architectural choices such as the number and width of layers and their type (e.g. dense, convolutional, recurrent), learning rate schedule, other parameters of the optimization scheme and regularizing mechanisms. Whereas for computer vision these successes were possible using raw pixels as the input representation, in the audio domain there seems to be an additional complication. Here the choices for how to best represent the input range from spectrograms, logarithmically filtered spectrograms over constant-Q transforms to even the raw audio itself \cite{dieleman_endtoend}.

This is a tedious problem, and there seem to be only two solutions to it: manual hyper-parameter selection, where a human expert tries to make decisions based on her past experience, or automatic hyper-parameter optimization as discussed in \cite{eggensperger_all_bayesho, bergstra_hyperopt, snoek_spearmint}. In this work we pursue a mixed strategy. As a first step, we systematically find the most suitable input representation, and progress from there with human expert knowledge to find best performing architectures for the task of framewise piano transcription.

A variety of neural network architectures has been used specifically for framewise transcription of piano notes from monaural sources. Some transcription systems are separated into an \textit{acoustic model} and a \textit{musical language model}, such as \cite{boulanger_temporal, sigtia_hybrid, sigtia_endtoend}, whereas in others there is no such distinction \cite{hanlon_polyphonic, boeck_polyphonic, bergkirkpatrick_unsupervised}. As shown in \cite{sigtia_endtoend}, models that utilize \textit{musical language models} perform better than those without, albeit the differences seem to be small. We focus on the \textit{acoustic model} here, neglecting the complementary language model for now.

\section{Input, Methods and Parameters}
\label{sec:methods}
In what follows, we will describe the input representations we compared, and give a brief overview of techniques for training and regularizing neural networks.

\subsection{Input Representation}
\label{sub:methods_input_representation}
Time-frequency representations in the form of spectrograms still seem to have a distinct advantage over the \textit{raw} audio input, as mentioned in \cite{dieleman_endtoend}. The exact parameterization of spectrograms is not entirely clear however, so we try to address this question in a systematic way.
We investigate the suitability of different types of spectrograms and constant-Q transforms as input representations for neural networks and compare four types of input representations: spectrograms with linearly spaced bins \textit{S}, spectrograms with logarithmically spaced bins \textit{LS}, spectrograms with logarithmically spaced bins and logarithmically scaled magnitude \textit{LM}, as well as the \mbox{constant-Q} transform \textit{CQT} \cite{brown_cqt}. The filterbank for \textit{LS} and \textit{LM} has a linear response (and lower resolution) for the lower frequencies, and a logarithmic response for the higher frequencies. We vary the sample rate $sr \in \{22050, 44100\}$ [Hz], number of bands per octave $nb \in \{12, 24, 36, 48\}$, whether or not frames undergo circular shift $cs \in \{\mathrm{on}, \mathrm{off}\}$, the amount of zero padding $zp \in \{\times 0, \times 1, \times 2\}$, and whether or not to use area normalized filters when filter banks are used $norm \in \{\mathrm{yes}, \mathrm{no}\}$. Furthermore, we re-scale the magnitudes of the spectrogram bins to be in the range $[0, 1]$. Table \ref{tab:grid_params} specifies which parameters are varied for which input type. For the computation of spectrograms we used Madmom \cite{madmom} and for the constant-Q transform we used the Yaafe library \cite{mathieu_yaafe}.

\begin{table}
\centering
\begin{tabular}{cccccc}
       & $sr$ & $zp$ & $cs$ & $nb$ & $norm$ \\
\textit{CQT}  & \x   &      &      & \x   &        \\
\textit{S}    & \x   & \x   & \x   &      &        \\
\textit{LS}   & \x   & \x   & \x   & \x   & \x     \\
\textit{LM}   & \x   & \x   & \x   & \x   & \x     \\
\end{tabular}
\caption{For each spectrogram type, these are the parameters that were varied. See text for a description of the value ranges.}
\label{tab:grid_params}
\end{table}

\subsection{Model Class and Suitability}
\label{subsec:model_class_suitability}
Formally, neural networks are functions with the structure

\begin{align*}
\mathrm{net}_{k}(\mathbf{x}) & = \mathbf{W}_k f_{k-1}(\mathbf{x}) + \mathbf{b}_{k} \\
f_{k}(\mathbf{x}) & = \sigma_{k}(\mathrm{net}_{k}(\mathbf{x})) \\
f_{0}(\mathbf{x}) & = \mathbf{x}
\end{align*}

\noindent where $\mathbf{x} \in \mathbb{R}^{w_{in}}$, $f_{k}: \mathbb{R}^{w_{k-1}} \rightarrow \mathbb{R}^{w_{k}}$, $\sigma$ is any element-wise nonlinear function, $\mathbf{W}_k$ is a matrix in $\mathbb{R}^{w_k \times w_{k-1}}$ called the \textit{weight matrix}, and $\mathbf{b}_{k} \in \mathbb{R}^{w_k}$ is a vector called \textit{bias}. The subscript $k \in \{0, \dots, L \}$ is the index of the layer, with $k=0$ denoting the input layer.

Choosing a very narrow definition on purpose, what we mean by a \textit{model class} $F$ is a fixed number of layers $L$, a fixed number of layer widths $\{w_0, \dots w_L\}$ and fixed types of nonlinearities $\{\sigma_0, \dots \sigma_L\}$. A \textit{model} means an instance $f$ from this class, defined by its weights alone. References to the whole collection of weights will be made with $\Theta$.

For the task of framewise piano transcription we define the \textit{suitability} of an input representation in terms of the performance of a simple classifier on this task, when given exactly this input representation.

Assuming we can reliably mitigate the risk of overfitting, we would like to argue that this method of determining suitable input representations, and using them for models with higher capacity, is the best we can do, given a limited computational budget.

Using a low-variance, high-bias model class, the perceptron, also called \textit{logistic regression} or \textit{single-layer} neural network, we learn a \textit{spectral template} per note. To test whether the results stemming from this analysis are really relevant for higher-variance, lower-bias model classes, we run the same set of experiments again, employing a multi layer perceptron with exactly one hidden layer, colloquially called a \textit{shallow net}. This small extension already gives the network the possibility to learn a shared, distributed representation of the input. As we will see, this has a considerable effect on how suitability is judged.

\subsection{Nonlinearities and Initialization}
Common choices for nonlinearities include the \textit{logistic function} $\sigma(a) = \frac{1}{1 + e^{-a}}$, \textit{hyperbolic tangent} $\sigma(a) = \tanh{a}$, and \textit{rectified linear units (ReLU)} $\sigma(a) = \max(0, a)$ . Nonlinearities are necessary to make neural networks universal function approximators \cite{hornik_universalapproximator}. According to \cite{he_initialization, glorot_initialization}, using ReLUs as the nonlinearities in neural networks leads to better behaved gradients and faster convergence because they do not saturate.

Before training, the weight matrices are initialized randomly. The scale of this initialization is crucial and depends on the used nonlinearity  as well as the number of weights contributing to the activation \cite{he_initialization, glorot_initialization}. Proper initialization plays an even bigger role when networks with more than one hidden layer are trained \cite{sussillo_randomwalk}. This is also important for the transcription setting we use, as the output layer of our networks uses the logistic function, which is prone to saturation effects. Thus we decided on using ReLUs throughout, initialized with a uniform distribution having a scale of $\pm \sqrt{2} \cdot \sqrt{\frac{2}{w_{k-1}+w_{k}}}$. For the last layer with the logistic nonlinearity, we omit the gain factor of $\sqrt{2}$, as advised in \cite{glorot_initialization}.

\subsection{Weight Decay}
To reduce overfitting and regularizing the network, different priors can be imposed on the network weights. Usually a Gaussian or Laplacian prior is chosen, corresponding to an $L_2$ or $L_1$ penalty term on connection weights, added to the cost function $\mathcal{L}_{reg} = \mathcal{L} + \lambda \sum_k \|\mathrm{vec}(\mathbf{W}_k)\|_{1|2}$ \cite{rumelhart_learning, williams_laplacian}, where $\mathcal{L}$ is an arbitrary, unregularized cost function and $\lambda$ governs the extent of regularization. Adding both of these penalty terms corresponds to a technique called \textit{Elastic Net} \cite{zou_elastic}. It is pointed out in \cite{bengio_practical} that using $L_2$ regularization plays a similar role as \textit{early stopping} and thus might be omitted. An $L_1$ penalty on the other hand leads to sparser weights, as it has a tendency to drive weights with irrelevant contributions to zero.

\subsection{Dropout}
Applying dropout to a layer zeroes out a fraction of the activations of a hidden layer of the network. For each training case, a different random fraction is dropped. This prevents units from co-adapting, and relying too much on each other's presence, as reasoned in \cite{srivastava_dropout}. Dropout increases robustness to noise, improves the generalization ability of networks and mitigates the risk of overfitting. Additionally dropout can be interpreted as model-averaging of exponentially many models \cite{srivastava_dropout}.

\subsection{Batch Normalization}
Batch normalization \cite{ioffe_batchnorm} seeks to produce networks whose individual activations per layer are zero-mean and unit-variance. This is ensured by normalizing the activations for each mini-batch at each training step. This effectively limits how far the activation distribution can drift away from zero-mean, unit-variance during training. Not only does this alleviate the need of the weights of the subsequent layer to adapt to a changing input distribution during training, it also keeps the nonlinearities from saturating and in turn speeds up training. It has additional regularizing effects, which become more apparent the more layers a network has. After training is stopped, the normalization is performed for each layer and for the whole training set.

\subsection{Layer Types}
We employ three different types of layer. Their respective functions can all be viewed as matrix expressions in the end, and thus can be made to fit into the framework described in Section \ref{subsec:model_class_suitability}. For the sake of readability, we simply describe their function in a procedural way.

A \textit{dense} layer consists of a dense matrix - vector pair $(\mathbf{W}, \mathbf{b})$ together with a nonlinearity. The input is transformed via this affine map, and then passed through a nonlinearity.

A \textit{convolutional} layer consists of a number $C_k$ of convolution kernels of a certain size $\{(\mathbf{W}_c, \mathbf{b}_c)\}_{c=0}^{C_k}$ together with a non-linearity. The input is convolved with each convolution kernel, leading to $C_k$ different feature maps to which the same nonlinearity is applied.
 
\textit{Max pooling} layers are used in convolutional networks to provide a small amount of translational invariance. They select the units with maximal activation in a local neighborhood $(w_t, w_f)$ in time and frequency in a feature map. This has beneficial effects, as it makes the transcriber invariant to small changes in tuning.

\textit{Global average pooling} layers are used in all-convolutional networks to compute the mean value of feature maps.

\subsection{Architectures}
There is a fundamental choice between a network with all dense layers, a network with all convolutional layers, and a mixed approach, where usually the convolutional layers are the first ones after the input layer followed by dense layers. Pooling layers, batch normalization and dropout application are interleaved. 
For all networks we have to choose the number of layers, how many hidden units per layer to use and when to interleave a regularization layer. For convolutional networks we have to choose the number of filter kernels and their extent in time and frequency direction.

\subsection{Networks for Framewise Polyphonic Piano Transcription}
The output layer of all considered model classes has $88$ units, in line with the playable notes on most modern pianos, and the output nonlinearity is the logistic function, whose output ranges lie in the interval $[0, 1]$.

The loss function being minimized is the frame- and element-wise applied \textit{binary crossentropy}
\begin{align*}
  \mathcal{L}_{bce}^{(t)}(\mathbf{y}_t, \hat{\mathbf{y}}_t) = -(\mathbf{y}_t \cdot \log(\hat{\mathbf{y}}_t) + (1 - \mathbf{y}_t) \cdot \log(1 - \hat{\mathbf{y}}_t))
\end{align*}

where $\hat{\mathbf{y}}_t = f_L(\mathbf{x}_t)$ is the output vector of the network, and $\mathbf{y}_t$ the ground truth at time $t$. As the overall loss over the whole training set we take the mean

\begin{align*}
\mathcal{L} = \frac{1}{T}\sum_{t=1}^{T} \mathcal{L}_{bce}^{(t)}
\end{align*}

For the purpose of computing the performance measures, the prediction of the network is thresholded to obtain a binary prediction $\bar{\mathbf{y}}_t = \hat{\mathbf{y}}_t > 0.5$.

\subsection{Optimization}
The simplest way to adapt the weights $\Theta$ of the network to minimize the loss is to take a small step with length $\alpha$ in the direction of steepest descent:

\begin{align*}
 \Theta_{i + 1} = \Theta_{i} - \alpha \cdot \frac{\partial \mathcal{L}}{\partial \Theta} 
\end{align*}

Computing the true gradient $\frac{\partial \mathcal{L}}{\partial \Theta} = \frac{1}{T} \sum_{t=1}^{T} \frac{\partial \mathcal{L}_{bce}^{(t)}}{\partial \Theta}$ requires a sum over the length of the whole training set, and is computationally too costly. For this reason, the gradient is usually only approximated from an i.i.d. random sample of size $M \ll T$.\ This is called \textit{mini-batch stochastic gradient descent}. There are several extensions to this general framework, such as \textit{momentum} \cite{polyak_momentum}, \textit{Nesterov momentum} \cite{nesterov_momentum} or \textit{Adam} \cite{kingma_adam}, which try to smooth the gradient estimate, correct small missteps or adapt the learning rate dynamically, respectively. Additionally we can set a \textit{learning rate schedule} that controls the temporal evolution of the learning rate.

\section{Dataset and Measures}
\label{sec:datasets_and_measures}
The computational experiments have been performed with the MAPS dataset \cite{emiya_multipitch}. It provides MIDI-aligned recordings of a variety of classical music pieces. They were rendered using different hi-quality piano sample patches, as well as real recordings from an upright Disklavier. This ensures clean annotation and therefore almost no label-noise. For all performance comparisons the following framewise measures on the validation set are used:

\newcommand{\TP}{\mathit{TP}}
\newcommand{\FP}{\mathit{FP}}
\newcommand{\FN}{\mathit{FN}}

\begin{align*}
P &= \sum_{t=0}^{T-1}\frac{\TP[t]}{\TP[t] + \FP[t]} \\
R &= \sum_{t=0}^{T-1}\frac{\TP[t]}{\TP[t] + \FN[t]} \\
F_1 &= \frac{2 \cdot P \cdot R}{P + R}
\end{align*}

The train-test folds are those used in \cite{sigtia_endtoend} which were published online \footnote{\url{http://www.eecs.qmul.ac.uk/~sss31/TASLP/info.html}}. For each fold, the validation set consists of $43$ tracks randomly removed from the train set, deviating from the $26$ used in \cite{sigtia_endtoend}, and leading to a division of 173-43-54 between the three sets. Note that the test sets are the \textit{same}, and are referred to as \textit{configuration I} in \cite{sigtia_endtoend}. The exact splits for \textit{configuration II} were not published. We had to choose them ourselves, using the same methodology, which has the additional constraint that \textit{only recordings of the real piano} are used for testing, resulting in a division of 180-30-60. This constitutes a more realistic setting for piano transcription.

\section{Analysis of relative Hyper-Parameter Importance}
To identify and select an appropriate input representation and determine the most important hyper-parameters responsible for high transcription performance, a multi-stage study with subsequent fANOVA analysis was conducted, as described in \cite{hutter_efficient}. This is similar in spirit to \cite{greff_lstm_space}, albeit on a smaller scale.

\subsection{Types of Representation}
To isolate the effects of different input representations on the performance of different model classes, only parameters for the spectrogram were varied according to Table \ref{tab:grid_params}. This leads to $204$ distinct input representations. The hyper-parameters for the model class as well as the optimization scheme were held \textit{fixed}. To make our estimates more robust, we conducted multiple runs for the same type of input.

The results for each model class are summarized in Table \ref{tab:model_classes_effects}, containing the three most influential hyper-parameters and the percentage of variability in performance they are responsible for. The most important hyper-parameter for both model classes is the type of spectrogram used, followed by pairwise interactions. Please note that the numbers in the percentage column are mainly useful to judge the \textit{relative} importance of the parameters. We will see these relative importances put into a larger context later on.

\begin{figure}[htp]
     \centering
     \begin{subfigure}{.49\linewidth}
       \includegraphics[scale=0.28]{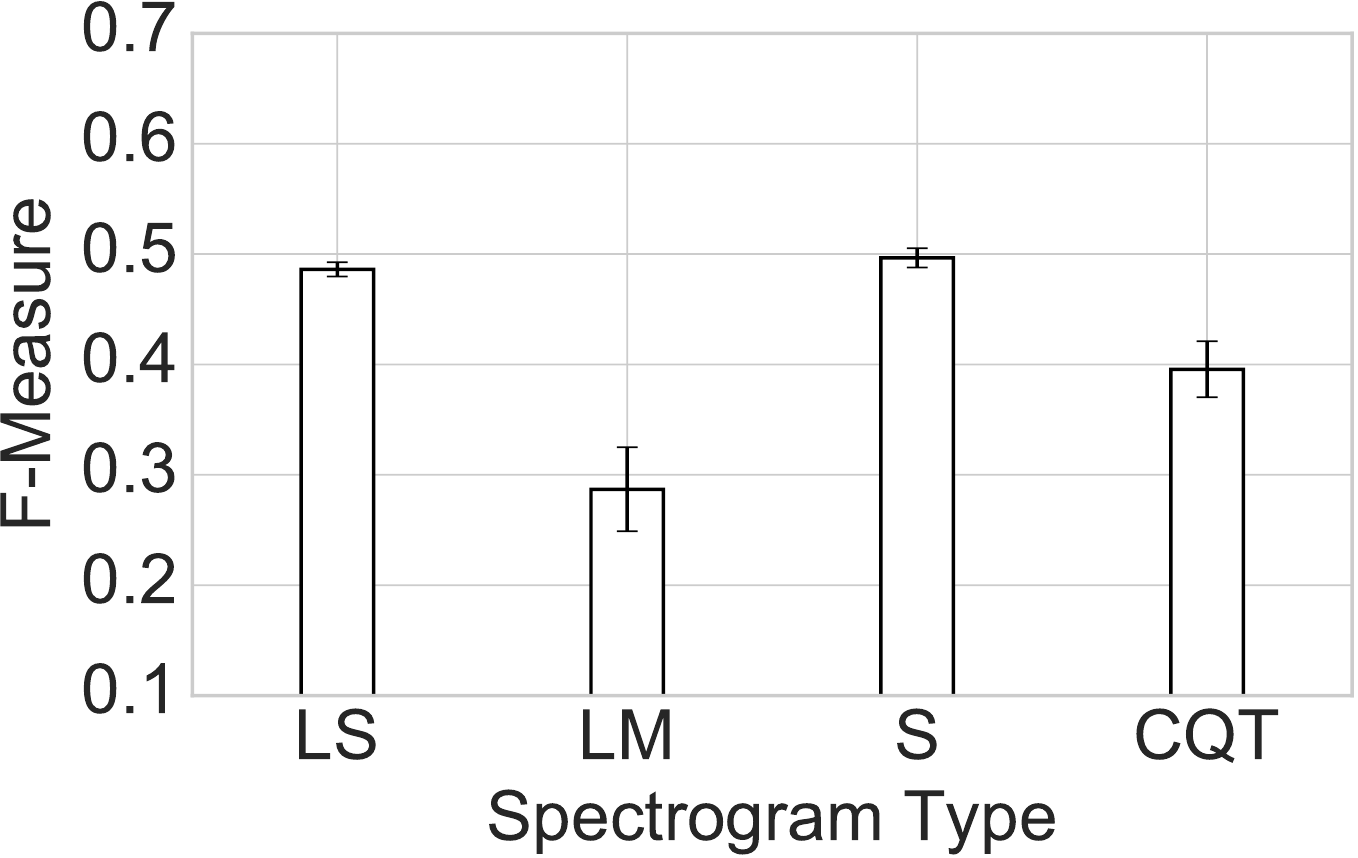}\hspace{-1em}
        \label{fig:input_main_effect_logreg}
     \end{subfigure}
     \begin{subfigure}{.49\linewidth}
       \includegraphics[scale=0.28]{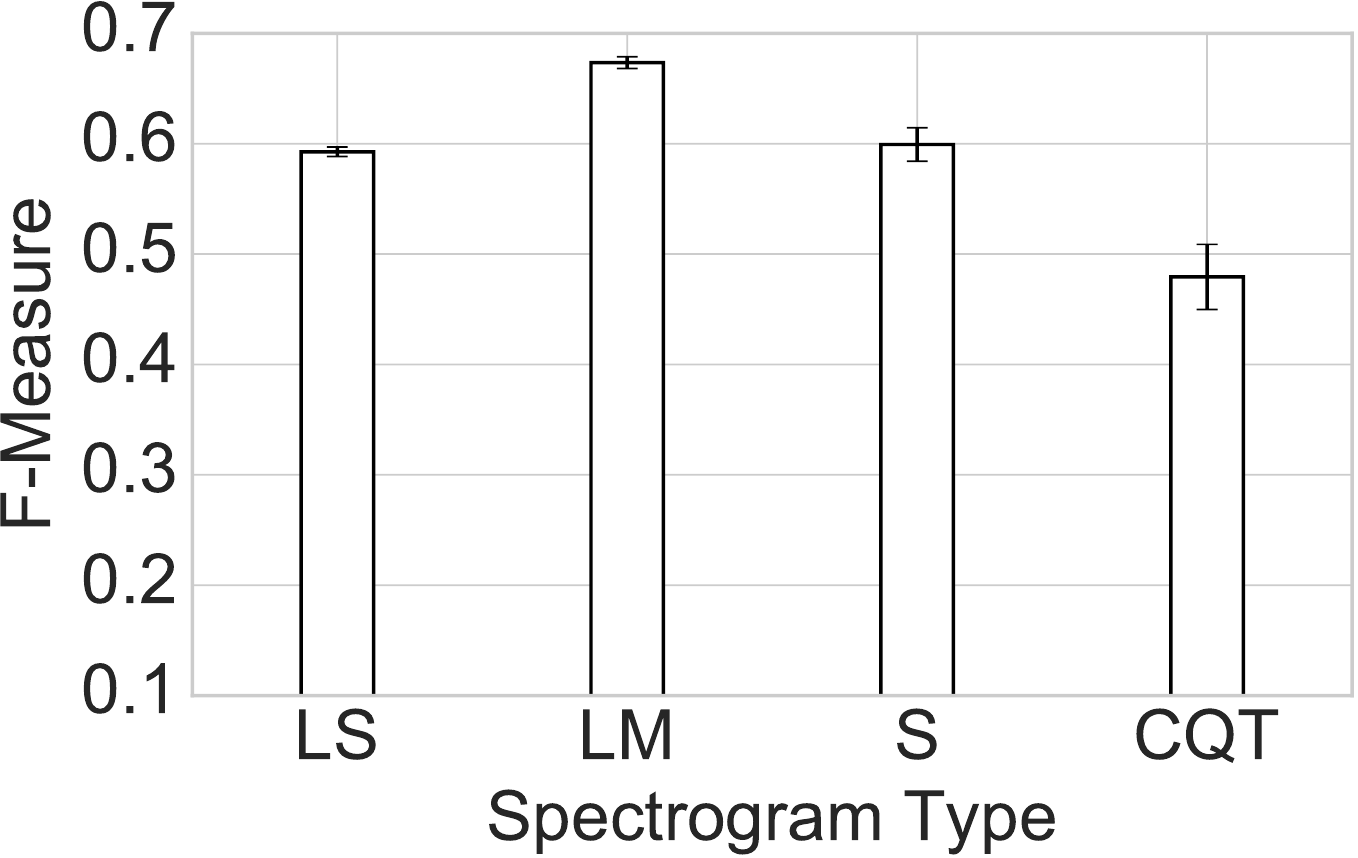}
       \label{fig:input_main_effect_shallow}
     \end{subfigure}
     \caption{(a) Mean logistic regression performance dependent on spectrogram (b) Mean shallow net performance dependent on type of spectrogram}
     \label{fig:input_main_effect}
\end{figure}

In Figure \ref{fig:input_main_effect}, we can see the mean performance attainable with different types of spectrograms for both model classes. The error bars indicate the standard deviation for the spread in performance, caused by the rest of the varied parameters. Surprisingly, the spectrogram with logarithmically spaced bins and logarithmically scaled magnitude, $LM$, enables the shallow net to perform best, even though it is a clear mismatch for logistic regression. The lower performance of the constant-Q transform was quite unexpected in both cases and warrants further investigation.

\begin{table}[htp]
\centering
\begin{tabular}{l|c|l}
Model Class         & Pct    & Parameters \\
\hline
\hline
Logistic Regression & 48.6\% & Spectrogram Type \\
                    & 16.9\% & Spectrogram Type \\
                    &        & $\times$ Normed Area Filters \\
                    & 10.4\% & Spectrogram Type \\
                    &        & $\times$ Sample Rate \\
\hline
Shallow Net         & 68.4\% & Spectrogram Type \\
                    & 20.8\% & Spectrogram Type \\
                    &        & $\times$ Sample Rate \\ 
                    & 5.7\%  & Sample Rate 
\end{tabular}
\caption{The three most important parameters determining input representation for different model classes}
\label{tab:model_classes_effects}
\end{table}

\subsection{Greater context}

Attempting a full grid search on all possible input representation and model class hyper-parameters described in Section \ref{sec:methods} to compute their true marginalized performance is computationally too costly. It is possible however to compute the \textit{predicted} marginalized performance of a hyper-parameter efficiently from a smaller subsample of the space, as shown in \cite{hutter_efficient}. All parameters are randomly varied to sample the space as evenly as possible, and a random forest of $100$ regression trees is fitted to the measured performance. This allows to \textit{predict} the marginalized performance of individual hyper-parameters. Table \ref{tab:additional_values} contains the list of hyper-parameters varied.

\begin{figure}[htp]
  \centering
  \includegraphics[scale=0.45]{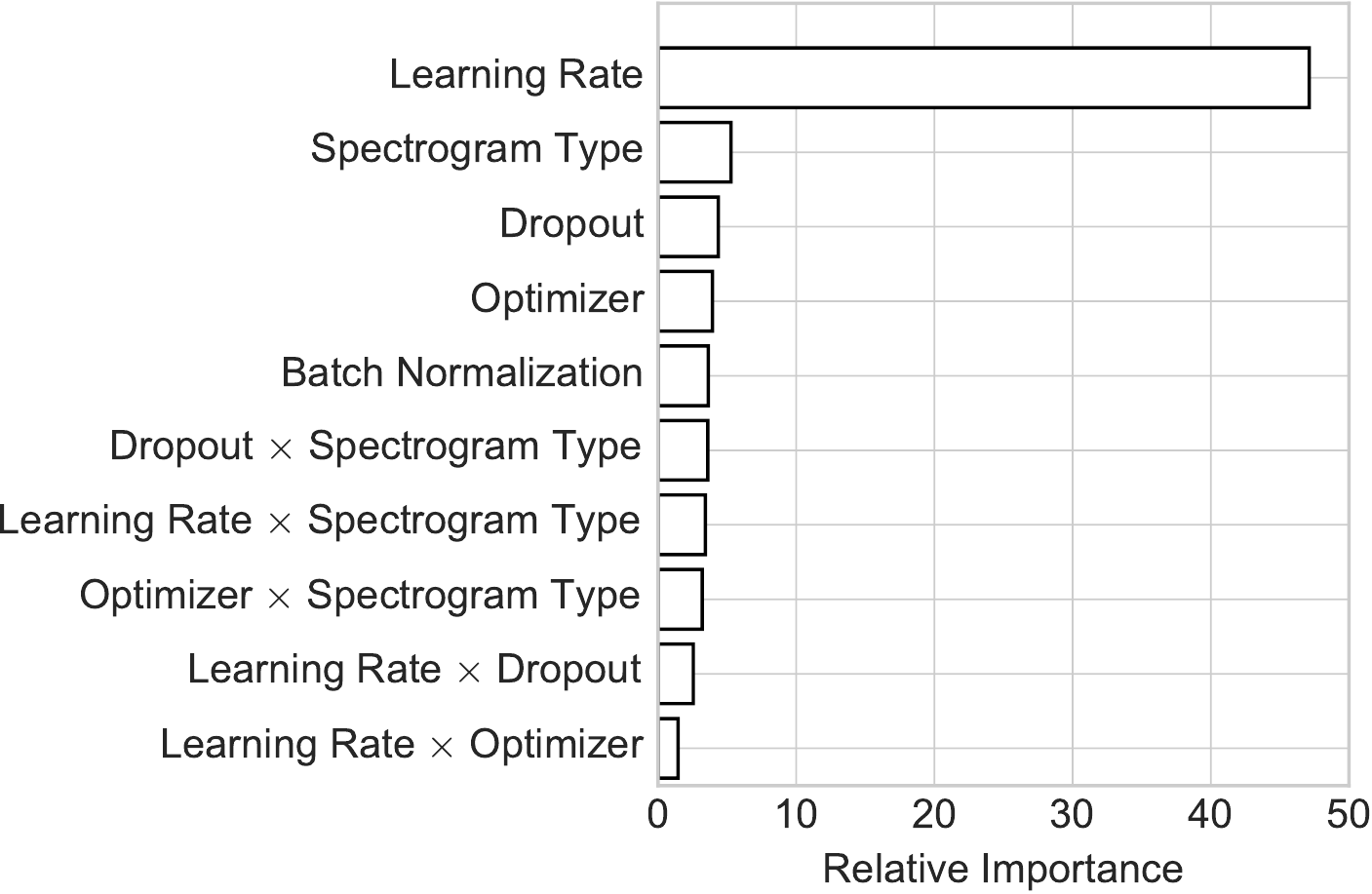}
     \caption{Relative importance of the first $10$ hyper-parameters for the \textit{shallow net} model class.}
     \label{fig:all_marginals}
\end{figure}

The percentage of variance in performance these hyper-parameters are responsible for, can be seen in Figure \ref{fig:all_marginals} for the $10$ most important ones. A total of $3000$ runs with random parameterizations were made.

\begin{table}[htp]
\centering
\small
\begin{tabular}{l}
Optimizer (Plain SGD, Momentum, Nesterov Momentum, Adam) \\
Learning Rate (0.001, 0.01, 0.1, 0.5, 1.0, 2.0, 10.0, 50.0, 100.0) \\
Momentum (Off, 0.7, 0.8, 0.9) \\
Learning rate Scheduler (On, Off) \\
Batch Normalization (On, Off) \\
Dropout (Off, 0.1, 0.3, 0.5) \\
$L_1$ Penalty (Off, 1e-07, 1e-08, 1e-09) \\
$L_2$ Penalty (Off, 1e-07, 1e-08, 1e-09)
\end{tabular}
\caption{The list of additional hyper-parameters varied, and their ranges.}
\label{tab:additional_values}
\end{table}

\begin{figure}[htp]
  \centering
  \includegraphics[scale=0.4]{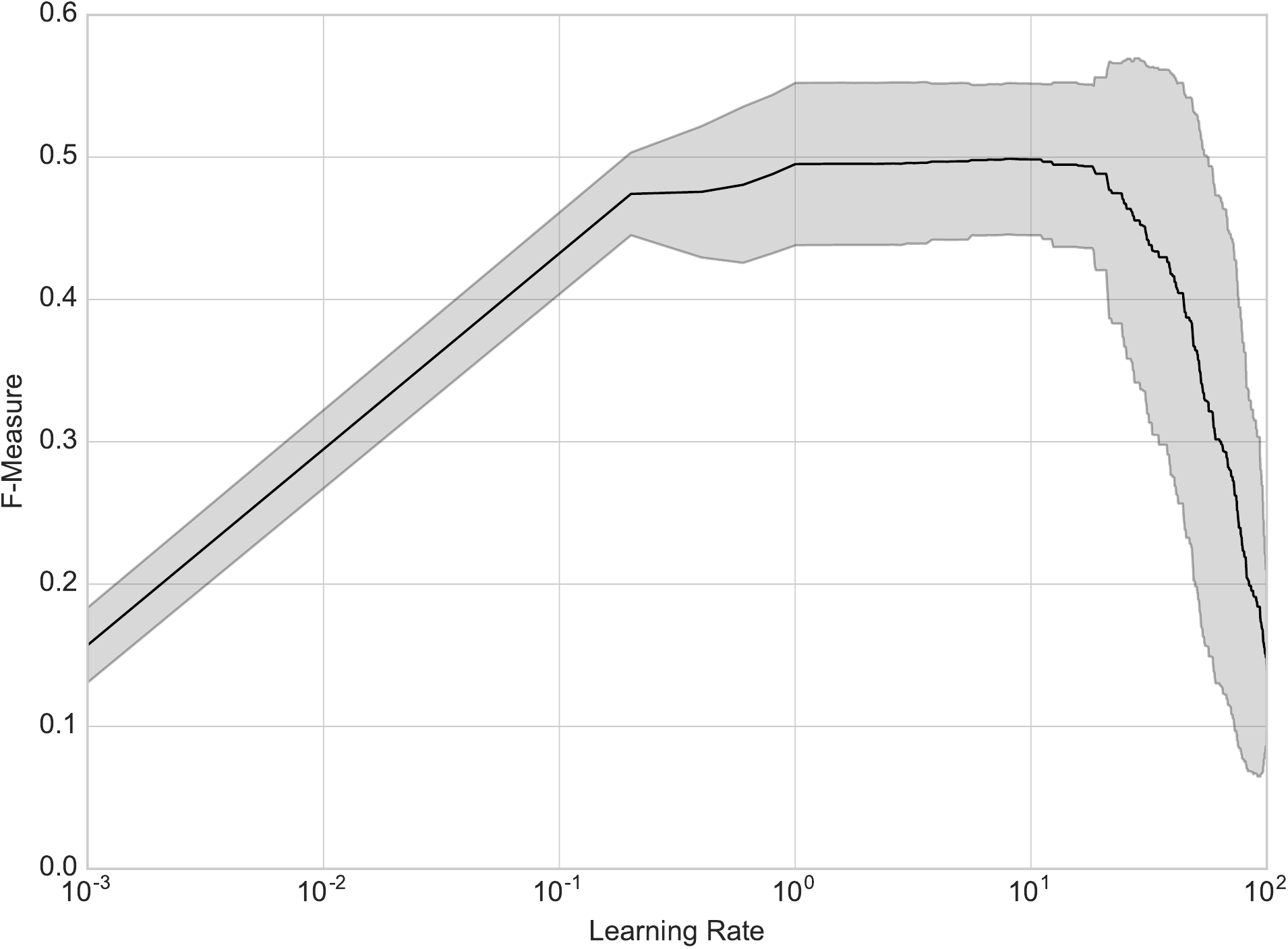}
     \caption{Mean predicted performance for the \textit{shallow net} model class, dependent on learning rate (on a logarithmic scale). The dark line shows the mean performance, and the gray area shows the standard deviation.}
     \label{fig:all_marginals_most_important}
\end{figure}

Analyzing the results of all the runs tells us that the most important hyper-parameters are \textit{Learning Rate} (47.11\%), and \textit{Spectrogram Type} (5.28\%). The importance of the learning rate is in line with the findings in \cite{greff_lstm_space}. Figure \ref{fig:all_marginals} shows the relative importances of the first $10$ hyper-parameters, and Figure \ref{fig:all_marginals_most_important} shows the predicted marginal performance of the learning rate dependent on its value (on a logarithmic scale) in greater detail. 

\section{State of the Art Models}
Having completed the analysis of input representation, more powerful model classes were tried: a deep neural network consisting entirely of dense layers (\textit{DNN}), a mixed network with convolutional layers directly after the input followed by dense layers (\textit{ConvNet}), and an all-convolutional network (\textit{AllConv} \cite{springenberg_allconv}). Their architectures are described in detail in Table \ref{tab:model_architectures}. To the best of our knowledge, this is the first time an all-convolutional net has been proposed for the task of framewise piano transcription.

We computed a logarithmically filtered spectrogram with logarithmic magnitude from audio with a sample rate of $44.1$ kHz, a filterbank with $48$ bins per octave, normed area filters, no circular shift and no zero padding. The choices for circular shift and zero padding  ranged very low on the importance scale, so we simply left them switched off. This resulted in only $229$ bins, which are logarithmically spaced in the higher frequency regions, and almost linearly spaced in the lower frequency regions as mentioned in Section \ref{sub:methods_input_representation}. The dense network was presented one frame at a time, whereas the convolutional network was given a context in time of two frames to either side of the current frame, summing to $5$ frames in total.

\begin{table}[htp]
\centering
\small
\begin{tabular}{l|l|l|l}
&\textit{DNN} & \textit{ConvNet} & \textit{AllConv} \\
\hline
\hline
&Input 229    & Input 5x229   &  Input 5x229 \\
&Dropout 0.1  & Conv 32x3x3   &  Conv 32x3x3 \\
&Dense 512    & Conv 32x3x3   &  Conv 32x3x3 \\
&BatchNorm    & BatchNorm     &  BatchNorm \\
&Dropout 0.25 & MaxPool 1x2   &  MaxPool 1x2 \\
&Dense 512    & Dropout 0.25  &  Dropout 0.25 \\
&BatchNorm    & Conv 64x3x3   &  Conv 32x1x3 \\
&Dropout 0.25 & MaxPool 1x2   &  BatchNorm \\
&Dense 512    & Dropout 0.25  &  Conv 32x1x3 \\
&BatchNorm    & Dense 512     &  BatchNorm \\
&Dropout 0.25 & Dropout 0.5   &  MaxPool 1x2 \\
&Dense 88     & Dense 88      &  Dropout 0.25 \\
&             &               &  Conv 64x1x25 \\
&             &               &  BatchNorm \\
&             &               &  Conv 128x1x25 \\
&             &               &  BatchNorm \\
&             &               &  Dropout 0.5 \\
&             &               &  Conv 88x1x1 \\
&             &               &  BatchNorm \\
&             &               &  AvgPool 1x6 \\
&             &               &  Sigmoid \\
\hline
\# Params & 691288       &  1877880      & 284544
\end{tabular}
\caption{Model Architectures}
\label{tab:model_architectures}
\end{table}

All further hyper-parameter tuning and architectural choices have been left to a human expert. Models within a model class were selected based on average F-measure across the four validation sets. An automatic search via a hyper-parameter search algorithm for these larger model classes, as described in \cite{eggensperger_all_bayesho, bergstra_hyperopt, snoek_spearmint} is left for future work (the training time for a convolutional model is roughly $8-9$ hours on a Tesla K40 GPU, which leaves us with $204 \cdot 3 \cdot 4 \cdot 8$ hours (variants $\times$ \#models $\times$ \#folds $\times$ hours per model), or on the order of $800 - 900$ days of compute time to determine the best input representation exactly).

For these powerful models, we followed practical recommendations for training neural networks via gradient descent found in \cite{bengio_practical}. Particularly relevant is the way of setting the initial learning rate. Strategies that dynamically adapt the learning rate, such as \textit{Adam} or \textit{Nesterov Momentum} \cite{kingma_adam, nesterov_momentum} help to a certain extent, but still do not spare us from tuning the initial learning rate and its schedule.

We observed that using a combination of batch normalization and dropout together with very simple optimization strategies leads to low validation error fairly quickly, in terms of the number of epochs trained. The strategy that worked best for determining the learning rate and its schedule was trying learning rates on a logarithmic scale, starting at $10.0$, until the optimization did not diverge anymore \cite{bengio_practical}, then training until the validation error flattened out for a few epochs, then multiplying the learning rate with a factor from the set $\{0.1, 0.25, 0.5, 0.75\}$. The rates and schedules we finally settled on were:

\begin{itemize}
\item \textit{DNN}: SGD with Momentum, $\alpha = 0.1, \mu = 0.9$ and halving of $\alpha$ every $10$ epochs
\item \textit{ConvNet}: SGD with Momentum, $\alpha = 0.1, \mu = 0.9$ and a halving of $\alpha$ every $5$ epochs
\item \textit{AllConv}: SGD with Momentum, $\alpha = 1.0, \mu = 0.9$ and a halving of $\alpha$ every $10$ epochs
\end{itemize}

The results for framewise prediction on the MAPS dataset can be found in Table \ref{tab:results_configuration_i}. It should be noted that we compare straightforward, simple, and largely un-smoothed systems (ours) with hybrid systems \cite{sigtia_endtoend}. There is a small degree of temporal smoothing happening when processing spectrograms with convolutional nets. The term \textit{simple} is supposed to mean that the resulting models have a small amount of parameters and the models are composed of a few low-complexity building blocks. All systems are evaluated on the same train-test splits, referred to as \textit{configuration I} in \cite{sigtia_endtoend} as well as on \textit{realistic} train-test splits, that were constructed in the same fashion as \textit{configuration II} in \cite{sigtia_endtoend}.

\begin{table}[htp]
\centering
\small
\begin{tabular}{l|l|l|l}
Model Class    & $P$ & $R$ & $F_1$ \\
\hline
\hline
Hybrid DNN \cite{sigtia_endtoend}     & 65.66 & 70.34 & 67.92 \\
Hybrid RNN \cite{sigtia_endtoend}     & 67.89 & 70.66 & 69.25 \\
Hybrid ConvNet \cite{sigtia_endtoend} & 72.45 & 76.56 & 74.45 \\
\hline
\textit{DNN}            & 76.63 & 70.12 & 73.11 \\
\textit{ConvNet}        & 80.19 & \textbf{78.66} & \textbf{79.33} \\
\textit{AllConv}        & \textbf{80.75} & 75.77 & 78.07 \\
\end{tabular}
\caption{Results on the MAPS dataset. Test set performance was averaged across $4$ folds as defined in \textit{configuration I} in \cite{sigtia_endtoend}.}
\label{tab:results_configuration_i}
\end{table}

\begin{table}[htp]
\centering
\small
\begin{tabular}{l|l|l|l}
Model Class    & $P$ & $R$ & $F_1$ \\
\hline
\hline
DNN \cite{sigtia_endtoend}     & - & - & 59.91 \\
RNN \cite{sigtia_endtoend}     & - & - & 57.67 \\
ConvNet \cite{sigtia_endtoend} & - & - & 64.14 \\
\hline
\textit{DNN}            & 75.51 & 57.30 & 65.15 \\
\textit{ConvNet}        & 74.50 & \textbf{67.10} & \textbf{70.60} \\
\textit{AllConv}        & \textbf{76.53} & 63.46 & 69.38 \\
\end{tabular}
\caption{Results on the MAPS dataset. Test set performance was averaged across $4$ folds as defined in \textit{configuration II} in \cite{sigtia_endtoend}.}
\label{tab:results_configuration_ii}
\end{table}

\section{Conclusion}
\label{sec:conclusion}
We argue that the results demonstrate: the importance of proper choice of input representation, and the importance of hyper-parameter tuning, especially the tuning of learning rate and its schedule; that convolutional networks have a distinct advantage over their deep and dense siblings, because of their context window and that all-convolutional networks perform nearly as well as mixed networks, although they have far fewer parameters. We propose these straightforward, framewise transcription networks as a new state-of-the art baseline for framewise piano transcription for the MAPS dataset.

\section{Acknowledgements}
This work is supported by the European Research Council (ERC Grant Agreement 670035, project \mbox{CON ESPRESSIONE}),
the Austrian Ministries BMVIT and BMWFW, the Province of Upper Austria (via the COMET Center SCCH)
and the European Union Seventh Framework Programme FP7 / 2007-2013 through the GiantSteps project (grant agreement no. 610591).
We would like to thank all developers of Theano \cite{theano} and Lasagne \cite{lasagne} for providing comprehensive and easy to use deep learning frameworks. The Tesla K40 used for this research was donated by the NVIDIA Corporation.

\bibliography{master}

\end{document}